\documentstyle[epsf,epsfig,11pt, here]{article}
\def\be{\begin{equation}}
\def\ee{\end{equation}}
\def\bea{\begin{eqnarray}}
\def\eea{\end{eqnarray}}

\newcommand{\rhobar}{\overline{\rho}}
\newcommand{\etabar}{\overline{\eta}}
\newcommand{\epsilonk}{\left|\epsilon_K \right|}
\newcommand{\vubovcb}{\left | \frac{V_{ub}}{V_{cb}} \right |}

\newcommand{\dmd}{\Delta m_d}
\newcommand{\dms}{\Delta m_s}

\newcommand{\Dstar}{{\rm D}^{\ast}}

\newcommand{\Bbar}{\overline{{\rm B}}}

\newcommand{\nubar}{\overline{\nu_{\ell}}}

\newcommand{\Vcb}{\left | {\rm V}_{cb} \right |}
\newcommand{\Vub}{\left | {\rm V}_{ub} \right |}

\newcommand{\mumu}{\ifmmode {\mu^+\mu^-} \else ${\mu^+\mu^-} $ \fi}

\newcommand{\ea}{\end{array}}
\newcommand{\bc}{\begin{center}}
\newcommand{\ec}{\end{center}}
\newcommand{\Kz}{\ifmmode {\rm K^0_s} \else ${\rm K^0_s} $ \fi}
\newcommand{\Zz}{\ifmmode {\rm Z^0} \else ${\rm Z^0 } $ \fi}
\newcommand{\qqbar}{\ifmmode {\rm q\bar{q}} \else ${\rm q\bar{q}} $ \fi}
\newcommand{\ccbar}{\ifmmode {\rm c\bar{c}} \else ${\rm c\bar{c}} $ \fi}
\newcommand{\bbbar}{\ifmmode {\rm b\bar{b}} \else ${\rm b\bar{b}} $ \fi}
\newcommand{\xxbar}{\ifmmode {\rm x\bar{x}} \else ${\rm x\bar{x}} $ \fi}
\newcommand{\rphi}{\ifmmode {\rm R\phi} \else ${\rm R\phi} $ \fi}
\newcommand{\bt}{\begin{tabular}}
\newcommand{\et}{\end{tabular}}

\begin{document}

\begin{flushright}
{\large\bf LAL 00-04}\\
\vspace*{0.1cm}
{\bf LPT-Orsay 00-20}\\
\vspace*{0.1cm}
{\large February 2000}
\end{flushright}
\vskip 1.0 cm

\title{\bf Determination of the CKM unitarity triangle parameters
by end 1999}

\begin {center}
{\Large\bf F. Caravaglios} \\
\vspace*{0.1cm}
{\large\bf Laboratoire de Physique Th\'eorique}, {\it CNRS et Universit\'e de Paris-Sud,\\
B\^at. 210 - 91405 Orsay Cedex}
\end{center}
\vspace*{0.1cm}
\begin {center}
{\Large\bf F. Parodi}\\
\vspace*{0.1cm}
{\large\bf Universit\'a di Genova and INFN}, {\it Dipartimento di Fisica \\
16126 Genova, Italy}
\end{center}
\vspace*{0.1cm}
\begin {center}
{\Large\bf P. Roudeau and A. Stocchi}\\
\vspace*{0.1cm}
{\large\bf Laboratoire de l'Acc\'el\'erateur Lin\'eaire}, {\it IN2P3-CNRS et Universit\'e de Paris-Sud\\
B\^at. 200 - B.P. 34 - 91898 Orsay Cedex}
\end{center}
\vspace*{1cm}
\begin{abstract}
Within the Standard Model, a review of the current determination 
of the CKM unitarity triangle parameters is presented, using experimental constraints 
from the measurements of $\epsilonk$, $\vubovcb$, $\dmd$ and from the limit on $\dms$, available by end 1999.
\end{abstract}
\vspace*{0.5cm}
\section{Introduction}
\label{sec:1}
This paper contains an update of the determination of the CKM parameters,
mainly $\rhobar$ and $\etabar$, of the Wolfenstein parametrization and of the angles of the unitarity triangle, using
experimental results and theoretical estimates available at the end of 1999.
Previous analyses using the same approach can be found in \cite{ref:cimento,ref:scripta}.
Other determinations of these parameters can be found in \cite{ref:autresharicot}. As compared to end 1998, 
new experimental results concern mainly first averages of LEP measurements of $\Vcb$ and $\Vub$ \cite{ref:steering} and 
an improved limit on $\dms$ \cite{ref:osciw}.
\section{Determination of the $CKM$ unitarity triangle parameters in the Standard Model framework}
\label{sec:2}
Results have been obtained using the constraints from the measurements of $\epsilonk$, $\vubovcb$, $\dmd$ and $\dms$.
The central values and the uncertainties for the relevant parameters used in this analysis are given in Table \ref{tab:1}.
As compared with the similar analysis done in \cite{ref:cimento}, the following changes have been made:

i) the uncertainty on $\Vcb$ \cite{ref:steering} has been increased from $1.5\times10^{-3}$
to $1.9\times10^{-3}$. This uncertainty is entirely dominated by theoretical 
errors and the current belief is that these errors amount to 5$\%$ \cite{ref:ligeti}.
 The quoted central value for $\Vcb$, in Table \ref{tab:1}, corresponds to the present LEP
average of $\Vcb$ measurements from inclusive and exclusive B semileptonic
decays. Correlations between theoretical errors explain why the quoted 
uncertainty happens to be slightly below 5$\%$. \\
ii) the values quoted for the ratio $\vubovcb$ have been obtained using
the value of $\Vcb$ mentioned before and the values of $\Vub$ measured
at CLEO with exclusive decays and at LEP from inclusive analyses.\\
iii) The limit on $\Delta m_s$, $\Delta m_s > 14.3~ps^{-1} ~{\rm at}~95\%\ {\rm C.L.}$ \cite{ref:osciw}, is higher than the one quoted
at the EPS-HP99 Conference after the addition of recent measurements from the SLD Collaboration. The sensitivity \cite{ref:osciw} of present measurements is equal to $14.5~ps^{-1}$.
\begin{table}[h]
\caption{\it Values of the most relevant quantities entering into the expressions of $\epsilonk$, $\vubovcb$, $\dmd$ and $\dms$.
In the third column the Gaussian and the flat part of the errors are given explicitly. }
\label{tab:1} 
\begin{center}
\vspace*{0.56cm}
\normalsize
\begin{tabular}{|c|c|c|c|}
\hline
&&&\\
                       Parameter                 &  Values                      & Gaussian $-$ Flat error   &  Ref. \\ 
&&&\\
\hline
&&&\\ 
                    $\lambda$                    & $0.2205 \pm 0.0018$          & $\pm 0.0018 - \pm 0.000$  & \cite{ref:cimento}  \\
$\left | V_{cb} \right |$                & $(40.5 \pm 1.9)\times 10^{-3}$   & $(\pm 1.9 - \pm 0.0)\times 10^{-3}$ &\cite{ref:steering}\\
$\frac{\left | V_{ub} \right |}{\left | V_{cb} \right |}$ (CLEO) & $0.085 \pm 0.018$ & $\pm 0.009 - \pm 0.016$ &\cite{ref:cleo}  \\
$\frac{\left | V_{ub} \right |}{\left | V_{cb} \right |}$ (LEP)  & $0.104 \pm 0.019$ & $\pm 0.011 - \pm 0.015$ &\cite{ref:steering} \\
                    $\Delta m_d$                 & $(0.476 \pm 0.016) ~ps^{-1}$ & $\pm 0.016 - \pm 0.000$   & \cite{ref:osciw}  \\
                    $\Delta m_s$     & $>$ 14.3 ps$^{-1}~{\rm at}~95\% ~C.L.$ & sensit. at 14.5 ps$^{-1}$  & \cite{ref:osciw}  \\
               $ \overline{m_t}(m_t)$            & $(167 \pm 5) ~GeV/c^2$       & $ \pm 5  - \pm 0$         & \cite{ref:topmass} \\
               $B_K$                & $0.86 \pm 0.10$               & $\pm 0.06  -  \pm 0.08$  & \cite{ref:cimento}     \\
               $ f_{B_d} \sqrt{B_{B_d}}$      & $(210 \pm 42)~MeV$ & $\pm 29  - \pm 31$ &\cite{ref:cimento}\\
$\xi=\frac{ f_{B_s}\sqrt{B_{B_s}}}{ f_{B_d}\sqrt{B_{B_d}}}$ & $1.11^{+0.06}_{-0.04}$ &$\pm 0.02 - ^{+0.06}_{-0.04}$&\cite{ref:cimento}\\
&&&\\
\hline 
\end{tabular}
\end{center}
\end{table}

\subsection{Values of $\rhobar$, $\etabar$ and of the angles of the unitarity triangle}
\label{sec:21}
\begin{figure}
\begin{center}
\vskip -0.5cm
{\epsfig{figure=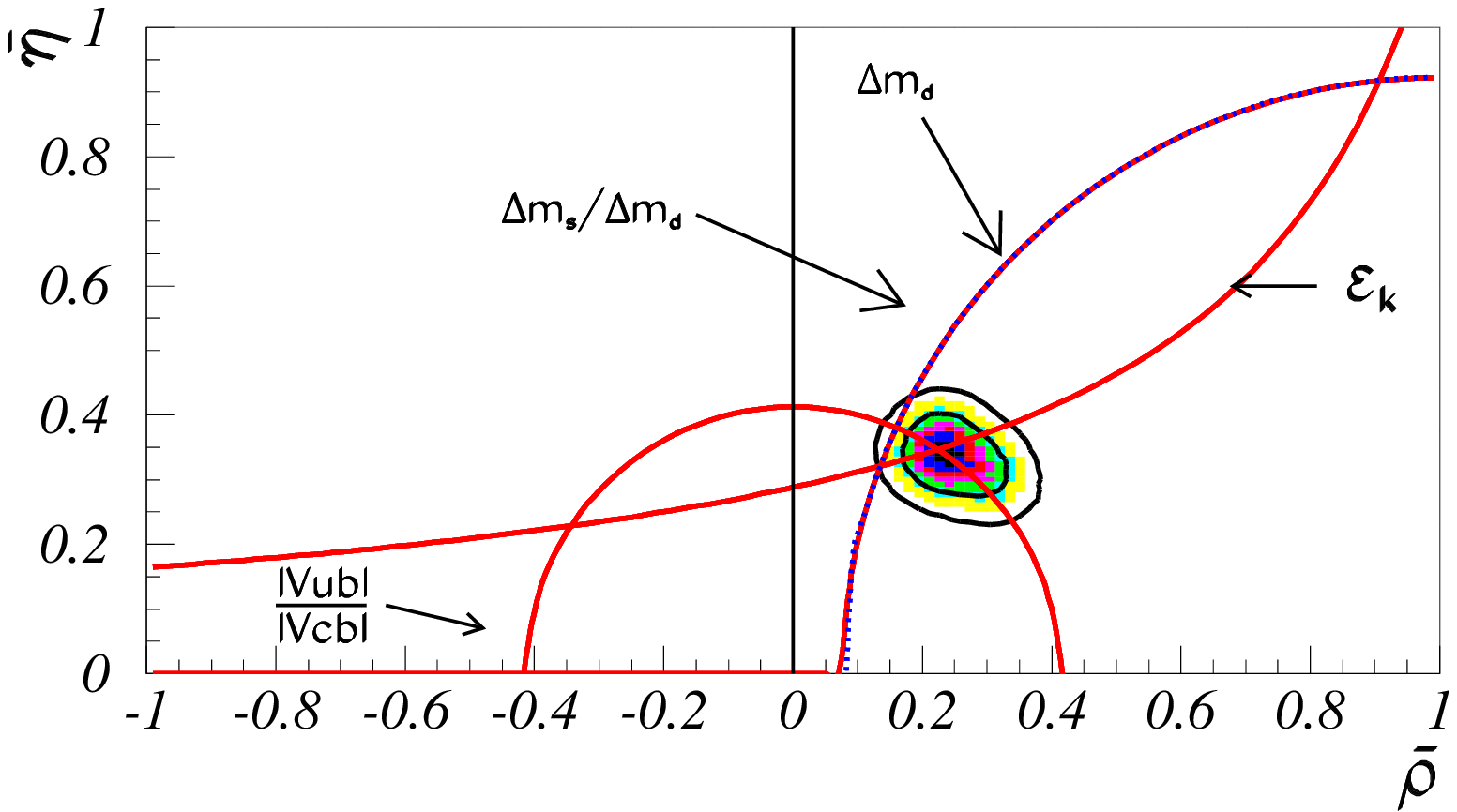,bbllx=30pt,bburx=503pt,bblly=1pt,bbury=270pt,height=7cm}}
\caption{ \it{ The allowed region for $\overline{\rho}$ and $\overline{\eta}$ using the parameters listed in Table \ref{tab:1}.
The contours at 68$\%$ and 95$\%$ are shown. The full lines correspond to the central values of the constraints given by
the measurements of  $  \frac{\left | V_{ub} \right |}{\left | V_{cb} \right |} $, $\epsilonk$ and $\Delta m_d$.
The dotted curve (which is hidden behind the curve of $\Delta m_d$) corresponds to the 95$\%$ C.L. upper limit 
obtained from the experimental limit on $\Delta m_s$.}}
\label{fig:rhoeta_bc99}
\vspace*{0.8cm}
{\epsfig{figure=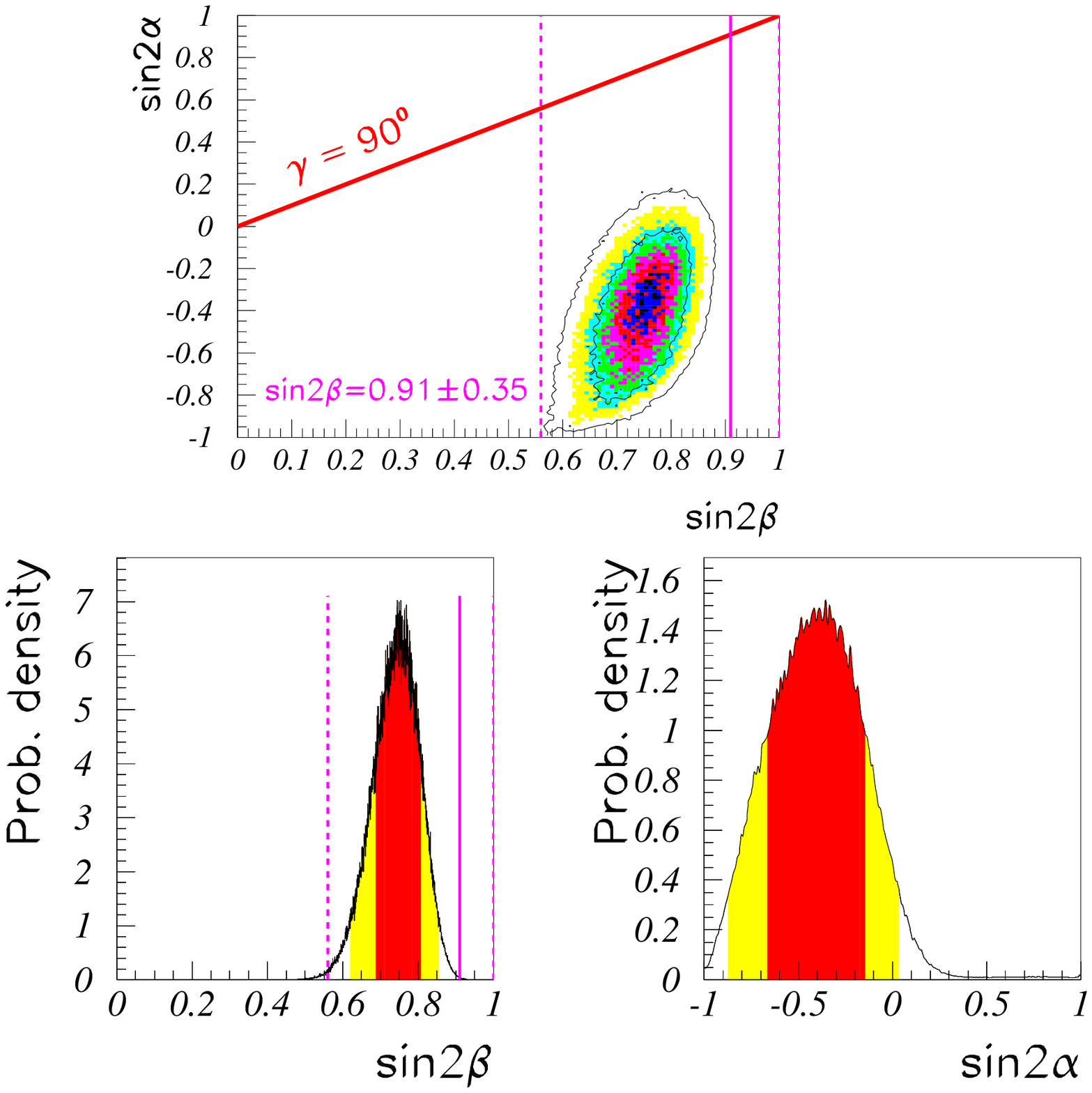,bbllx=0pt,bburx=560pt,bblly=0pt,bbury=515pt,height=11cm}}
\vskip -0.2cm
\caption{ \it{ The $sin 2 \alpha$ and $sin 2 \beta$ distributions have been obtained using the constraints corresponding
to the values of the parameters listed in Table \ref{tab:1}.  The dark-shaded and the clear-shaded intervals correspond,
respectively, to 68$\%$ and 95$\%$ confidence level regions. The lines corresponding to the $\pm 1 \sigma$ errors from 
the measurement of sin 2$\beta$ using $J/ \Psi,K^0_s$ events are also shown. The line obtained for $\gamma = 90^{\circ}$ 
is also drawn (sin2$\beta$ = sin2$\alpha$). }}
\label{fig:seni_bc99}
\end{center}
\end{figure}

The region in the $(\rhobar,~\etabar)$ plane selected by the measurements of $\epsilonk$,
$\vubovcb$, $\dmd$ and from the limit on $\dms$, is given in Figure \ref{fig:rhoeta_bc99}.
The measured values of the two parameters are:
\begin{equation}
\rhobar=0.240^{+0.057}_{-0.047}  ,~\etabar=0.335 \pm 0.042 
\label{eq:rhoeta2}
\end{equation}
Fitted values for the angles of the unitarity triangle have been also obtained:
\begin{equation}
\sin(2\beta)=  0.750^{+0.058}_{-0.064}   ,~\sin(2\alpha)= -0.38^{+0.24}_{-0.28}~{\rm and}~\gamma=(55.5^{+6.0}_{-8.5})^{\circ}.
\label{eq:angles}
\end{equation}

\begin{figure}[h]
\begin{center}
{\epsfig{figure=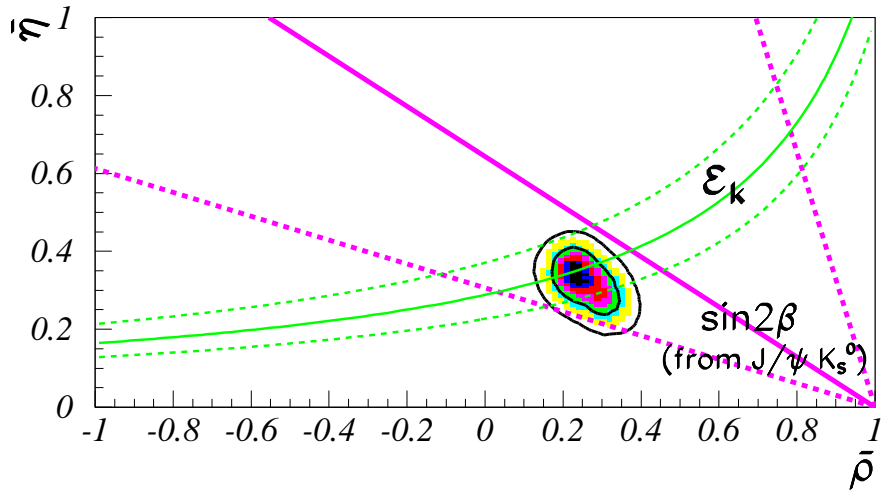,bbllx=30pt,bburx=503pt,bblly=1pt,
bbury=270pt,height=7cm}}
\caption{\it The allowed region for $\rhobar$ and $\etabar$ using the parameters listed in Table \ref{tab:1}.
The contours at 68$\%$ and 95$\%$ are shown. The constraint from $\epsilonk$ has not been used and the dotted lines 
correspond to 68$\%$ C.L.. The lines corresponding to the $\pm 1 \sigma$ errors from the measurement of 
sin 2$\beta$ using $J/ \Psi,K^0_s$ events are also shown (sin2$\beta$=0.91$\pm$0.35) \cite{ref:forty}.}
\label{fig:rhoeta_sinepsk_bc99}
\end{center}
\end{figure}

The accuracy on sin2$\beta$ obtained with the present analysis, is better than the one expected from B factories 
but it is valid only in the Standard Model framework. 
First measurements of $sin(2\beta)$ are now available. 
The world average is:  $sin(2\beta)$ = 0.91 $\pm$ 0.35 \cite{ref:forty}. The 68$\%$ C.L. region is shown in Figure 
\ref{fig:seni_bc99} and the constraints on $\rhobar$ and $\etabar$ from this measurement are shown in 
Figure \ref{fig:rhoeta_sinepsk_bc99}. \\
The angle $\gamma$ is known with a 15$\%$ relative error. Values of $\gamma > $90$^{\circ}$ 
(or $\rhobar <$ 0) are excluded at 99.6\% C.L.. 
The origin of asymmetric errors on $\rhobar$, $\gamma$ and sin2$\beta$ is that the information brought by the limit on $\Delta m_s$
is quite efficient in constraining values of $\Delta m_s$ up to 15 ps$^{-1}$ but no information can be obtained for values above 
20 ps$^{-1}$.

\newpage

\noindent
The values of the non-perturbative QCD parameters can be also obtained:
\begin{equation}
f_B\sqrt{B_B} = (232 \pm 13) MeV    ~~~~~ ; ~~~~~ B_K = 0.80 ^{+0.15}_{-0.17}
\label{eq:fb}
\end{equation}
after having removed, in turn, these constraints from the analysis. It can be noticed that $f_B\sqrt{B_B}$ is better 
determined than the present evaluation of this parameter from lattice QCD calculations (Table \ref{tab:1}).
As a consequence, it is important to observe, contrarily to common belief, that a large uncertainty attached to 
$f_B\sqrt{B_B}$ has no real impact on the present analysis. Evaluation of this parameter with 5-10$\%$ relative error 
is needed to bring additional information. On the contrary, the parameter $B_K$ is determined with a 20$\%$ 
relative error using the other constraints. The present estimate of this parameter from lattice QCD calculations, given 
in Table \ref{tab:1}, has thus an  impact on the present analysis.\\
A region of the $(\rhobar,~\etabar)$ plane can be selected without using the $\epsilonk$  constraint \cite{ref:barbieri}. 
The result is shown in Figure~\ref{fig:rhoeta_sinepsk_bc99}. 
This test shows that the $(\rhobar,~\etabar)$  region selected by the measurements in the B sector (of the two 
sides of the unitarity triangle) is very well compatible with the region selected from the measurement of
CP violation in the kaon sector. Using B decay and oscillations properties only, the values of $\etabar$=0.325$\pm$ 0.054 
and $sin 2\beta$=0.747$^{+0.067}_{-0.084}$ 
are determined. 
\section{Often asked questions}
\label{sec:3}

We would like to dedicate this section to two often asked questions. \\
{\it {Question 1: The results from this analysis are precise since, both from the constraints and for
the parameters, Gaussian distributions have been taken for the experimental and the theoretical errors.}} \\
The answer is that we do not use only Gaussian distributions ! Values for the different parameters entering into 
the equations of constraints are extracted using random generations from Gaussian/non Gaussian distributions
depending on the source of the error (Table \ref{tab:1}). This has been shown at several seminars and conferences 
and explained in detail in \cite{ref:cimento}. For any parameter or constraint, entering into the fitting procedure, a detailed 
study has been done \cite{ref:cimento,ref:scripta}.   \\
{\it {Question 2: How the results are affected if theoretical errors are multiplied by a factor 2 ?}}

We believe that the aim of this work is to try to use at best the available measurements 
and theoretical estimates. Nevertheless it is a simple basic exercise to quantify the dependence of the uncertainties on the 
final results from the variation of the errors attached to the different quantities used in the analysis. 
This was already done in \cite{ref:scripta,ref:cimento}.
The exercise presented here consists in multiplying by a factor two all the ranges (flat distributions) used for the 
theoretical estimates given in Table \ref{tab:1} and taking for $V_{cb}$ an error of $\pm 3.0~10^{-3}$. 
Multiplying theoretical errors by a factor two seems to be rather extreme. In some cases (as for $V_{cb}$ or some 
lattice QCD parameters) this is equivalent to ignore more than 10 years of theoretical efforts. Just two examples. 
For the extraction of $V_{cb}$ using exclusive decays $\Bbar \rightarrow \Dstar \ell^- \nubar$, the conservative 
value of $F_{D^{\ast}}(w=1)$ = 0.90 $\pm$ 0.05 has been already taken (in \cite{ref:xxx} the suggested value is 0.913$\pm$0.042). 
In the following exercise the error of $\pm$ 0.10 is used ! In this exercise the theoretical 
error on $f_B \sqrt B_B$, coming from the evaluation of the quenched approximation, is taken to be $\pm$ 62 MeV. 
This type of errors circulated within the Lattice QCD community in the late '80. Table \ref{tab:2} 
summarizes the results of this exercise.
As a general conclusion, if theoretical errors on all the parameters given in Table \ref{tab:1} are simultaneously 
multiplied by a factor 2, the errors on the values of  $\rhobar$, $\etabar$, sin2$\beta$, sin2$\alpha$ and $\gamma$ 
increase by only a factor 1.5.

\begin{table}[h]
\caption{\it Errors on $\rhobar$, $\etabar$, sin2$\beta$, sin2$\alpha$ and $\gamma$ obtained after multiplying by a factor 2 the
flat part of theoretical errors given in Table \ref{tab:1} for the parameters mentioned in the first column. Inside brackets, 
values corresponding to the 95$\%$ C.L. are also given.}
\label{tab:2} 
\begin{center}
\vspace*{0.5cm}
\frame{\normalsize {
\begin{tabular}{c|c|c|c}
&&&\\
             Parameter     &  $\rhobar$        & $\etabar$  & sin2$\beta$  \\
&&&\\
\hline
&&&\\
 Standard 
& 0.240$^{+0.057}_{-0.047}$(0.150-0.335) & 0.335 $\pm$ 0.042($\times$ 2) & 0.750$^{+0.058}_{-0.064}$(0.618-0.858) \\
$f_B \sqrt{B_B} $ 
& 0.245$^{+0.060}_{-0.045}$(0.157-0.364) & 0.334 $\pm$ 0.042($\times$ 2) & 0.752$^{+0.058}_{-0.064}$(0.621-0.860) \\
$B_K $            
& 0.240$^{+0.058}_{-0.047}$(0.153-0.356) & 0.334 $\pm$ 0.044($\times$ 2) & 0.753$^{+0.058}_{-0.064}$(0.620-0.860) \\
$\xi $            
& 0.244$^{+0.060}_{-0.059}$(0.129-0.356) & 0.338 $\pm$ 0.043($\times$ 2) & 0.750$^{+0.056}_{-0.067}$(0.622-0.857)  \\
$\frac{\left | V_{ub} \right |}{\left | V_{cb} \right |}  $  
& 0.251$^{+0.067}_{-0.056}$(0.115-0.386) & 0.345 $\pm$ 0.050($\times$ 2) & 0.784$^{+0.077}_{-0.089}$(0.591-0.916)  \\
$\left | V_{cb} \right | $  
& 0.243$^{+0.066}_{-0.049}$(0.158-0.373) & 0.322 $\pm$ 0.049($\times$ 2) & 0.747$^{+0.059}_{-0.078}$(0.582-0.860)  \\
All             
& 0.278 $\pm$ 0.079 & 0.334 $\pm$ 0.061($\times$ 2) & 0.800$^{+0.073}_{-0.120}$(0.575-0.919)\\
&&&\\
\hline
\hline
&&&\\
             Parameter     &  sin2$\alpha$        & $\gamma$  & $\gamma <$90$^{\circ}$ \\
&&&\\
\hline
&&&\\
 Standard  
& -0.38$^{+0.24}_{-0.28}$(-0.89-0.06) & (55.5$^{+6.0}_{-8.5})^{\circ}$(38.4-67.3)$^{\circ}$ &  99.6$\%$ C.L.\\
$f_B \sqrt{B_B} $  
& -0.41$^{+0.24}_{-0.30}$(-0.91-0.04) & (55.1$^{+5.7}_{-9.5}$)$^{\circ}$(37.3-66.4)$^{\circ}$ & 99.6$\%$ C.L. \\
$B_K $          
& -0.37$^{+0.25}_{-0.29}$(-0.91-0.07) & (55.8$^{+6.1}_{-9.3}$)$^{\circ}$(37.8-67.1)$^{\circ}$  & 99.6$\%$ C.L. \\
$\xi $          
&-0.41$\pm$0.30 ($\times$ 2) & (54.5$\pm$ 8.3)$^{\circ}$($\times 2$)  & 99.6$\%$ C.L.  \\
$\frac{\left | V_{ub} \right |}{\left | V_{cb} \right |} $  
& -0.37$^{+0.25}_{-0.29}$(-0.91-0.09)& (55.0$^{+6.0}_{-8.8}$)$^{\circ}$(38.2-66.7)$^{\circ}$ & 99.5$\%$ C.L.  \\
$\left | V_{cb} \right | $                                  
&-0.45$^{+0.27}_{-0.35}$(-1.0-0.0) & (54.3$^{+6.9}_{-10.7}$)$^{\circ}$(33.6-67.1)$^{\circ}$ & 99.5$\%$ C.L.  \\
All                                                       
&-0.54$^{+0.35}_{-0.33}$(-1.0-0.14) & (50.7$\pm$ 10.3)$^{\circ}$($\times$ 2) & 99.4$\%$ C.L. \\
&&&\\
\hline
\end{tabular}}}
\end{center}
\end{table} 

\section{Conclusions}
\begin{figure}[h]
\begin{center}
{\epsfig{figure=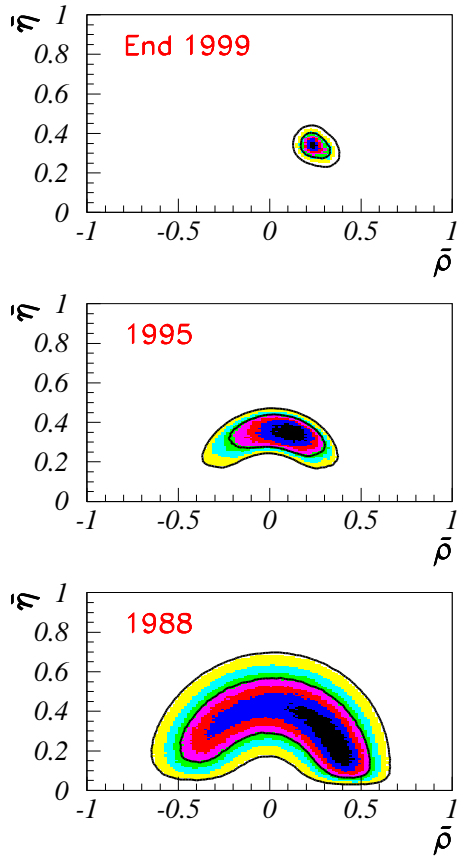,bbllx=30pt,bburx=350pt,bblly=1pt,
bbury=540pt,height=11.5cm}}
\caption{ \it{ The allowed region for $\overline{\rho}$ and $\overline{\eta}$ from 1988 to the end of 1999}}
\label{fig:rhoetastory_bc99}
\end{center}
\end{figure}

Impressive improvements have been accomplished during the last ten years in B physics. Figure \ref{fig:rhoetastory_bc99} 
illustrates the progress on the measurements of the two sides of the Unitarity Triangle. Some conclusions can be drawn.
The selected region in the $(\rhobar,\etabar)$ plane, using B physics only, is very well compatible with 
the measurement of CP violation in the Kaon system.
sin2$\beta$ is measured with an accuracy better than 10$\%$ within the SM framework. The angle $\gamma$ is smaller than 
90$^{\circ}$ at 99.6$\%$ C.L.. This result is very slightly affected by multypling theoretical errors 
by a factor two and is essentially due to the impressive improvement on the limit on $\Delta m_s$ obtained during the last 4 years.\\
The situation will still improve by summer 2000. Thanks to the achieved accuracy, future measurements of CP violation in 
the B sector would further test the consistency of the Standard Model by determining directly the angles of the Unitarity Triangle.

\section*{Acknowledgements}
Many thanks to the organizers for arranging such a nice Conference in a very stimulating atmosphere. The speaker 
has benefited from useful discussions with R. Forty, T. Nakada, R. Peccei, S. Willocq and M. Zito; they are warmly thanked. 

\end{document}